\documentclass[10pt]{article}
\usepackage{bbm}
\usepackage[final]{graphics}
\usepackage{amsmath}
\usepackage{amsfonts,amsbsy}
\usepackage{amssymb}
\usepackage[dvips]{color}

\def\empile#1\over#2{\mathrel{\mathop{\kern 0pt#1}\limits_{#2}}}
\def\bs{\boldsymbol}

\newcommand{\slv}{\raise.15ex\hbox{$/$}\kern-.53em\hbox{$v$}}
\newcommand{\slF}{\raise.15ex\hbox{$/$}\kern-.53em\hbox{$F$}}
\newcommand{\slL}{\raise.15ex\hbox{$/$}\kern-.53em\hbox{$L$}}
\newcommand{\slP}{\raise.15ex\hbox{$/$}\kern-.53em\hbox{$P$}}
\newcommand{\slp}{\raise.15ex\hbox{$/$}\kern-.53em\hbox{$p$}}
\newcommand{\slq}{\raise.15ex\hbox{$/$}\kern-.53em\hbox{$q$}}
\newcommand{\slR}{\raise.15ex\hbox{$/$}\kern-.53em\hbox{$R$}}
\newcommand{\slQ}{\raise.15ex\hbox{$/$}\kern-.53em\hbox{$Q$}}
\newcommand{\slK}{\raise.15ex\hbox{$/$}\kern-.53em\hbox{$K$}}
\newcommand{\slk}{\raise.15ex\hbox{$/$}\kern-.53em\hbox{$k$}}
\newcommand{\slD}{\raise.15ex\hbox{$/$}\kern-.53em\hbox{$D$}}
\newcommand{\slC}{\raise.15ex\hbox{$/$}\kern-.53em\hbox{$C$}}
\newcommand{\slA}{\raise.15ex\hbox{$/$}\kern-.53em\hbox{$A$}}
\newcommand{\slSigma}{\raise.15ex\hbox{$/$}\kern-.53em\hbox{$\Sigma$}}
\newcommand{\slpartial}{\raise.15ex\hbox{$/$}\kern-.53em\hbox{$\partial$}}
\newcommand{\slcalP}{\raise.15ex\hbox{$/$}\kern-.63em\hbox{$\cal P$}}

\def\p{{\boldsymbol p}}

\def\k{{\boldsymbol k}}

\def\x{{\boldsymbol x}}
\def\y{{\boldsymbol y}}

\catcode`\@=11

\newcount\@tempcntc
\def\@citex[#1]#2{\if@filesw\immediate\write\@auxout{\string\citation{#2}}\fi
  \@tempcnta\z@\@tempcntb\m@ne\def\@citea{}\@cite{%
        \@for\@citeb:=#2\do%
    {\@ifundefined{b@\@citeb}%
        {\@citeo\@tempcntb\m@ne\@citea%
                \def\@citea{,\penalty\@m\ }{\bf ?}\@warning%
                {Citation `\@citeb' on page \thepage \space undefined}}%
        {\setbox\z@\hbox{\global\@tempcntc0\csname b@\@citeb\endcsname\relax}
     \ifnum\@tempcntc=\z@ \@citeo\@tempcntb\m@ne%
       \@citea\def\@citea{,\penalty\@m}%
       \hbox{\csname b@\@citeb\endcsname}%
     \else%
      \advance\@tempcntb\@ne%
      \ifnum\@tempcntb=\@tempcntc%
      \else\advance\@tempcntb\m@ne\@citeo%
      \@tempcnta\@tempcntc\@tempcntb\@tempcntc\fi\fi}}\@citeo}{#1}}%

\def\@citeo{\ifnum\@tempcnta>\@tempcntb\else\@citea
  \def\@citea{,\penalty\@m}%
  \ifnum\@tempcnta=\@tempcntb\the\@tempcnta\else
   {\advance\@tempcnta\@ne\ifnum\@tempcnta=\@tempcntb \else
\def\@citea{--}\fi
    \advance\@tempcnta\m@ne\the\@tempcnta\@citea\the\@tempcntb}\fi\fi}

\catcode`\@=12

\begin{document}

\title{\bf Gluon propagation\\ inside a high-energy nucleus}
\author{Fran\c cois Gelis$^{(1)}$, Yacine Mehtar-Tani$^{(2)}$}
\maketitle
\begin{center}
\begin{enumerate}
\item Service de Physique Th\'eorique\footnote{URA 2306 du CNRS.}\\
  CEA/DSM/Saclay, B\^at. 774\\
  91191, Gif-sur-Yvette Cedex, France
\item Laboratoire de Physique Th\'eorique\\
  Universit\'e Paris Sud, B\^at. 210\\
  91405, Orsay cedex
\end{enumerate}
\end{center}

\begin{abstract}

       We show that, in the light-cone gauge, it is possible to derive
       in a very simple way the solution of the classical Yang-Mills
       equations for the collision between a nucleus and a proton.
       One important step of the calculation is the derivation of a
       formula that describes the propagation of a gluon in the
       background color field of the nucleus.  This allows us to
       calculate observables in pA collisions in a more
       straightforward fashion than already proposed. We discuss also
       the comparison between light-cone gauge and covariant gauge in
       view of further investigations involving higher order
       corrections.
\end{abstract}
\vskip 5mm
\begin{flushright}
Preprint SPhT-T05/194\\
LPT-Orsay 05-83
\end{flushright}

\section{Introduction}
The study of semi-hard particle production in high energy hadronic
interaction is dominated by interactions between partons having a
small fraction $x$ of the longitudinal momentum of the colliding
nucleons.  Since the phase-space density of such partons in the
nucleon wave function is large, one expects that the physics of parton
saturation \cite{GriboLR1,MuellQ1,BlaizM1} plays an important role in
such studies. This saturation generally has the effect of reducing the
number of produced particles compared to what one would have predicted
on the basis of pQDC calculation with parton densities that depend on
$x$ according to the linear BFKL \cite{BalitL1,KuraeLF1} evolution
equation.

It was proposed by McLerran and Venugopalan
\cite{McLerV1,McLerV2,McLerV3} that one could take advantage of this
large phase-space density in order to describe the small $x$ partons
by a classical color field rather than as particles. More precisely,
the McLerran-Venugopalan (MV) model proposes a dual description, in
which the small $x$ partons are described as a classical field and the
large $x$ partons act as color sources for the classical field. In
their original model, they had in mind a large nucleus, for which
there would be a large number of large $x$ partons (at least $3A$
where $A$ is the atomic number of the nucleus, from just counting the
valence quarks) and therefore they would produce a strong color
source. This meant that one has to solve the full classical Yang-Mills
equations in order to find the classical field. But that procedure, on
the other hand, would properly incorporate the recombination
interactions that are responsible for gluon saturation. In the MV
model, the large $x$ color sources are described by a statistical
distribution, which they argued could be taken to be a gaussian for a
large nucleus at moderately small $x$ (see also \cite{JeonV1} for a
more modern perspective on that).

Since then, this model has evolved into a full fledged effective
theory, the so-called ``Color Glass Condensate'' (CGC)
\cite{IancuLM1,IancuLM2,FerreILM1}. It was soon recognized that the
separation between what one calls large $x$ and small $x$, inherent to
the dual description of the MV model, is somewhat arbitrary, and that the
gaussian nature of the distribution of sources would not survive upon
changes of this separation scale. This arbitrariness has been
exploited to derive a renormalization group equation, the so-called
JIMWLK equation \cite{JalilKLW1,JalilKLW2,JalilKLW3,JalilKLW4,KovneM1,KovneMW3,JalilKMW1,IancuLM1,IancuLM2,FerreILM1}, that describes how
the statistical distribution of color sources changes as one moves the
boundary between large $x$ and small $x$. This functional evolution
equation can also be expressed as an infinite hierarchy of evolution
equations for correlators \cite{Balit1}, and has a quite useful (and
tremendously simpler) large $N_c$ mean-field approximation
\cite{Kovch3}, known as the Balitsky-Kovchegov equation.

In high energy hadronic collisions, gluon production is dominated by
the classical field approximation, and calculating it requires to
solve the classical Yang-Mills equations for two color sources moving
at the speed of light in opposite directions. This is a problem that
has been solved numerically in
\cite{KrasnV1,KrasnV2,KrasnNV1,KrasnNV2,Lappi1} for the
boost-invariant case, and the stability of this solution against
rapidity dependent perturbations has been investigated in
\cite{RomatV1,RomatV2}. In terms of analytical solutions, much less is
known. The only situation for which gluon production has been
calculated analytically is the case where one of the two sources is
weak and can thus be treated at lowest order (this situation is often
referred to as ``proton-nucleus'' collisions in the literature, but it
can also be encountered at forward rapidities in the collision of two
identical objects). This was done in a number of approaches
\cite{KovchM3,KovchT1,DumitM1,BlaizGV1}. The last two references
provide the solution of Yang-Mills equation in this asymmetrical
situation, in the Schwinger gauge ($x^+A^-+x^-A^+=0$) and Lorenz gauge
($\partial_\mu A^\mu=0$) respectively. More recently, Balitsky has
proposed an expansion in commutators of Wilson lines, where at each
order one treats the two projectiles symmetrically \cite{Balit2}.

Although the solution in the Lorenz gauge was fairly compact, it
turned out that it displayed some oddities like the appearance of a
Wilson line containing the coupling constant $g/2$ instead of $g$
(this of course disappeared at later stages from physical
quantities). When applied to the production of quark-antiquark
production in \cite{BlaizGV2}, it also led to a contribution in which
the vertex producing the $q\bar{q}$ pair was located inside the
nucleus, which is quite counter-intuitive.

In this paper, we derive the solution of the Yang-Mills equations in
the light-cone gauge $A^+=0$ (with the nucleus moving in the negative
$z$ direction), and we find a much simpler solution (not only the
solution is simpler, but it is also much easier to obtain). In
particular, the solution in the $A^+=0$ gauge presents none of the odd
features encountered in the Lorenz gauge. Our central result, derived
in section \ref{sec:prop}, is in fact a formula that tells how a color
field propagates on top of the gauge field of the nucleus. The
eqs.~(\ref{eq:LC-prop}) in fact contain all the information which is
needed in order to derive the solution of Yang-Mills equations and
compute gluon production, which we perform as a verification in
section \ref{sec:gluon}. Beyond the study of ``proton-nucleus''
collisions themselves, eqs.~(\ref{eq:LC-prop}) are an important
building block for calculating higher order corrections in the weak
source to the solution of Yang-Mills equations. And to a large extent,
the complexity of this object finds its way into the solution of
Yang-Mills equations. Therefore, it is important to determine this
object, and to find a gauge in which it is particularly simple. For
the sake of comparison, we derive in appendix \ref{sec:lorenz} the
analogue of eqs.~(\ref{eq:LC-prop}) for the Lorenz gauge, and they are
much more complicated, as one expects.

\section{Gluon propagation in a nucleus}
\label{sec:prop}
We assume that the nucleus is moving with a velocity very close to the
speed of light, in the negative $z$ direction. Thus, it can be
described by the following density of color sources:
\begin{equation}
\delta(x^+)\rho_a(\x_\perp)\; .
\end{equation}
Some intermediate calculations may require to regularize the delta
function by giving it a small width. When this is necessary, we
replace $\delta(x^+)$ by $\delta_\epsilon(x^+)$, where
$\delta_\epsilon(x^+)$ is a positive definite function normalized by
$\int_{-\infty}^{+\infty}dx^+\;\delta_\epsilon(x^+)=1$ and whose
support is $[0,\epsilon]$.

\begin{figure}[ht]
\begin{center}
\resizebox*{!}{2.5cm}{\includegraphics{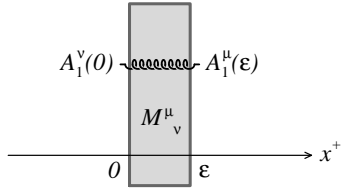}}
\caption{\label{Fig1}Gluon passing through a nucleus. The region
shaded in gray is the region where the color source representing the
nucleus is non-zero.}
\end{center}
\end{figure}

In this paper, we address the following question: knowing the
gauge fields at $x^+=0$, what are the gauge fields at
$x^+=\epsilon$, i.e. after having propagated through the nucleus?
Finding the gluon propagator inside the nucleus amounts to solving
this problem to linear order in the incoming gauge fields. In
practice, we add to the gauge field $A^\mu_0$ of the nucleus a
small perturbation $A^\mu_1$,
\begin{equation}
A^\mu=A_0^\mu+A_1^\mu\; ,
\end{equation}
and we wish to find a linear relationship for this perturbation before
and after the region where the nucleus lives, such as (see figure
\ref{Fig1})~:
\begin{equation}
A^\mu_1(x^+=\epsilon) = M^\mu{}_\nu \;A_1^\nu(x^+=0)\; .
\end{equation}

We work in the light-cone gauge $A^+=0$. The covariant conservation of
the color current reads\footnote{In this and in the following
equations, we do not write explicitly the color indices.}:
\begin{equation}
\partial^+J^- +[D^-,J^+]-[D^i,J^i]=0\; .
\end{equation}
This equation can be solved by
\begin{eqnarray}
&&J^+=J^i=0\; ,\nonumber\\
&&\partial^+J^-=0\; .
\end{eqnarray}
The first equation is allowed because all the color charges in our
problem are moving in the negative $z$ direction. The second equation
means that the color current $J^-$ associated to the nucleus is not
affected by the incoming field\footnote{In other gauges, the incoming
$A^+_1$ would induce a color precession of $J^-$.}, and therefore this
current $J^-$ can be ``hidden'' once for all in the gauge field
$A^\mu_0$ of the nucleus.

The Yang-Mills equations in this gauge read
\begin{eqnarray}
&&
\partial^+(\partial_\mu A^\mu)+ig[A^i,\partial^+A^i]=0
\nonumber\\
&&
[D^-,\partial^+A^-]-[D^i,F^{i-}]=J^-
\nonumber\\
&&
\partial^+F^{-i}+[D^-,\partial^+A^i]-[D^j,F^{ji}]=0\; .
\end{eqnarray}
One can see that the first of these equations does not contain any
time derivative ($\partial^-$). Therefore, it can be seen as a
constraint that relates the various field components at the same time.

The gauge field of the nucleus alone is found by considering only the
order zero in $A_1^\mu$ in the above equations. One can readily check
that the following is a solution:
\begin{equation}
A_0^i=0\quad,\quad
A_0^-=-\frac{1}{{\bs\partial}_\perp^2} J^-=-g\delta(x^+)
\frac{1}{{\bs\partial}_\perp^2}\rho(x_\perp)\; .
\end{equation}

In order to find the the linear relationship between the perturbation
$A_1^\mu$ before and after the region where the color sources of the
nucleus live, we must linearize the Yang-Mills equations in
$A_1^\mu$. One obtains
\begin{eqnarray}
&&
\partial^+ A_1^--\partial^i A_1^i=0
\nonumber\\
&&
\square A_1^i-2ig[A_0^-,\partial^+A^i]=0
\nonumber\\
&&
\square A_1^--2ig[A_0^-,\partial^+A_1^-]=2ig[\partial^iA_0^-,A_1^i]
=2ig (\partial^i A_0^-\cdot T)A_1^i\; .
\end{eqnarray}
The method for solving the second and third equations was explained in
\cite{BlaizGV1}. The solution of the second equation inside the
nucleus, i.e. for $x^+\in[0,\epsilon]$, reads:
\begin{equation}
A_1^i(x^+,x^-,\x_\perp)
=
U(x^+,0,\x_\perp) A_1^i(0,x^-,\x_\perp)\; ,
\end{equation}
where $U$ is a Wilson line in the adjoint representation of the
gauge group:
\begin{equation}
U(x^+,0,\x_\perp)\equiv {\cal T}_+\, \exp \left[ig\int_0^{x^+}
dz^+\; A_{0a}^-(z^+,\x_\perp)T^a\right]\; .
\end{equation}
At this point, we could find $A_1^-$ simply by solving the constraint
equation. But as a verification, it is interesting to solve explicitly
the third equation for $A_1^-$, and at the end to verify that the
fields $A_1^{i}$ and $A_1^-$ do obey the constraint. The third
equation leads to
\begin{eqnarray}
&&A_1^-(x^+,x^-,\x_\perp)
=
U(x^+,0,\x_\perp) A_1^-(0,x^-,\x_\perp)
\nonumber\\
&&
+
ig\int_0^{x^+}\!\!\!\!
dz^+ U(x^+,z^+,\x_\perp)(\partial^i A_0^-(z^+,\x_\perp)\cdot T)
U(z^+,0,\x_\perp)\frac{1}{\partial^+} A_1^i(0,x^-,\x_\perp)
\nonumber\\
&&=U(x^+,0,\x_\perp) A_1^-(0,x^-,\x_\perp)
+
(\partial^i U(x^+,0,\x_\perp))\frac{1}{\partial^+} A_1^i(0,x^-,\x_\perp)\; .
\end{eqnarray}
It is trivial to verify that the constraint that relates $A_1^-$ to
$A_1^i$ would have given the same answer.

Therefore, if we denote $U\equiv U(\epsilon,0,\x_\perp)$ and if we
don't write explicitly the $x^-$ and $\x_\perp$ dependence, we have
the following relations:
\begin{eqnarray}
&&
A_1^+(\epsilon)=0\; ,
\nonumber\\
&&
A_1^i(\epsilon)=U A_1^i(0)\; ,
\nonumber\\
&&
A_1^-(\epsilon)=U A_1^-(0)+(\partial^i U)\frac{1}{\partial^+}A_1^i(0)\; .
\label{eq:LC-prop}
\end{eqnarray}
These equations are the light-cone gauge expression of the linear
relation we were looking for. Analogous relations will be derived for
the Lorenz gauge in the appendix \ref{sec:lorenz}.

\section{Gluon production in pA collisions}
\label{sec:gluon}
\subsection{Gauge field}
\begin{figure}[ht]
\begin{center}
\resizebox*{!}{2.5cm}{\includegraphics{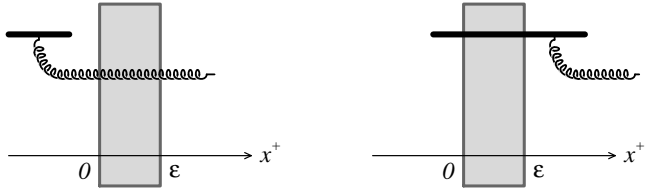}}
\caption{\label{Fig2} The two contributions to gluon production in pA
collisions. Left: the proton source emits the gluon before the
collision with the nucleus. Right: the proton source goes through the
nucleus before emitting the gluon. The thick solid line represents the
proton color current.}
\end{center} 
\end{figure}

>From this linear relation, it is easy to calculate the gauge field
that describes proton-nucleus collisions. In this case, the incoming
field $A_1^\mu$ is the field produced by the current associated to the
proton
\begin{equation}
J^+=g\delta(x^-)\rho_p(\x_\perp)\; .
\end{equation}
For $x^+\le 0$, i.e. before the collision with the nucleus, the
current $J^+$ remains constant, and we simply have, to linear order in
the proton source $\rho_p$
\begin{eqnarray}
&&
A_1^+=A_1^-=0\; ,
\nonumber\\
&&
\square A_1^i = -\theta(x^-)\partial^i \rho_p(\x_\perp)\; .
\end{eqnarray}
Solving the latter equation gives the following field at $x^+=0$~:
\begin{equation}
A_1^+(0)=A_1^-(0)=0\quad,\quad
A_1^i(0)
=\theta(x^-)\frac{\partial^i}{{\bs\partial}_\perp^2}\rho_p(\x_\perp)\; .
\label{eq:A1-0m}
\end{equation}
The next step is to use the eqs.~(\ref{eq:LC-prop}) in order to find
the gauge field $A_1^\mu$ immediately after the collision with the
nucleus, i.e. at $x^+=\epsilon$,
\begin{eqnarray}
&&
A_1^+(\epsilon)=0\; ,
\nonumber\\
&&
A_1^i(\epsilon)=
U\theta(x^-)\frac{\partial^i}{{\bs\partial}_\perp^2}\rho_p(\x_\perp)\; ,
\nonumber\\
&&
A_1^-(\epsilon)=
(\partial^i U)\frac{1}{\partial^+}\theta(x^-)
\frac{\partial^i}{{\bs\partial}_\perp^2}\rho_p(\x_\perp)\; .
\label{eq:field-eps}
\end{eqnarray}
The final step is to find the gauge field for $x^+>\epsilon$. The
equation that governs the evolution of $A_1^i$ is
\begin{equation}
\square A_1^i -2ig[A_0^-,\partial^+A_1^i]=-\frac{\partial^i}{\partial^+}J^+\; ,
\label{eq:EOM}
\end{equation}
and the current $J^+$ at $x^+>\epsilon$ is modified by color
precession in the nuclear field $A_0^-$:
\begin{equation}
J^+(x^+>\epsilon)=U \delta(x^-)\rho_p(\x_\perp)\; .
\end{equation}
This is a direct consequence of current conservation $D^-J^+=0$.  We
must solve eq.~(\ref{eq:EOM}) with an initial condition at
$x^+=\epsilon$ given by eq.~(\ref{eq:field-eps}). Using the techniques of
\cite{BlaizGV1}, we obtain
\begin{eqnarray}
&&
A_1^i(x)=\int_{y^+=\epsilon} dy^-d^2\y_\perp\;
G_{_R}^0(x,y)2\partial_y^+ U(\y_\perp) \theta(y^-)
\frac{\partial_y^i}{{\bs\partial}_{\y_\perp}^2}\rho_p(\y_\perp)
\nonumber\\
&&\quad+
\int_{y^+>\epsilon}d^4y\; G_{_R}^0(x,y) \theta(y^-)
\partial_y^i(U(\y_\perp)\rho_p(\y_\perp))\; ,
\label{eq:A1-0p}
\end{eqnarray}
where $G_{_R}^0(x,y)$ is the free retarded propagator obeying
$\square_x G_{_R}^0(x,y)=\delta(x-y)$. The two terms in this solution
are illustrated in figure \ref{Fig2}.

\subsection{Gluon production}
The amplitude for the production of a gluon of momentum $p$ and
polarization $\lambda$ is given by the Fourier transform of the
amputated gauge field, contracted in the relevant polarization
vector,
\begin{equation}
{\cal M}_\lambda(p)=\int d^4x \; e^{ip\cdot x} \square A_1^\mu(x)
\epsilon^{(\lambda)}_\mu(\p)\; .
\label{eq:ampl}
\end{equation}
In light-cone gauge, the sum over the physical polarizations reads
\begin{equation}
\sum_{\lambda}\epsilon^{(\lambda)}_i(\p)\epsilon^{(\lambda)*}_j(\p)=-g_{ij}\;
,
\end{equation}
and for this reason we need only the transverse components of the
gauge field when calculating gluon production. Note that since the
Dalembertian $\square A_1^i(x)$ is bounded inside the nucleus, the
region $0<x^+<\epsilon $ does not contribute in the limit $\epsilon\to
0$. Therefore, we can take $\epsilon\to 0$ and disregard the interior
of the nucleus when we calculate the amplitude. We first
obtain\footnote{When we evaluate the Dalembertian of the gauge field
in the region $x^+<0$, we must multiply eq.~(\ref{eq:A1-0m}) by
$\theta(-x^+)$ in order to restrict this term to the region
$x^+<0$. Equivalently, we could simply subtract it from
eq.~(\ref{eq:A1-0p}) in order not to overcount it in the region
$x^+>0$.}:
\begin{eqnarray}
\square A_1^i(x)&=&
2\delta(x^+)\delta(x^-)(U-1)\frac{\partial^i}{{\bs\partial}^2_\perp}
\rho_p(\x_\perp)
\nonumber\\
&-&
\theta(x^-)\theta(-x^+) \partial^i \rho_p(\x_\perp)
-\theta(x^-)\theta(x^+) \partial^i(U\rho_p(\x_\perp))\; ,
\end{eqnarray}
and then the Fourier transform gives
\begin{eqnarray}
-p^2 A_1^i(p)&=&
-p^2 A_{\rm proton}^i(p)
\nonumber\\
&+&
i\int \frac{d^2\k_{1\perp}}{(2\pi)^2}
\left[\frac{p^i}{(p^++i\varepsilon)(p^-+i\varepsilon)}
-\frac{k_1^i}{k_{1\perp}^2}
\right]
\nonumber\\
&& \qquad\qquad\times \rho_p(\k_{1\perp})
\left[U(\k_{2\perp})-(2\pi)^2\delta(\k_{2\perp})\right]\; .
\end{eqnarray}
In this equation, $\k_{2\perp}\equiv \p_\perp-\k_{1\perp}$ and
$A_{\rm proton}^i(p)$ is the Fourier transform of the gauge field
of a proton alone, i.e. the Fourier transform of
eq.~(\ref{eq:A1-0m}). It is easy to verify that this expression
leads to the standard result for gluon production in
proton-nucleus collisions\footnote{ In \cite{KovchM3,KovnW}, the same gauge was used but in a diagrammatic approach which deals with quark and gluon interactions with the nucleus, so that, the derivation of gluon production is more involved. 
}.

\section{Conclusions}
In this paper, we have obtained the solution of the classical
Yang-Mills equations for proton-nucleus collisions in the light-cone
gauge $A^+=0$. An important intermediate step is the ``transfer
matrix'', given in eqs.~(\ref{eq:LC-prop}), which tells how a color
field propagates through the nucleus on top of the color field of the
nucleus. It turns out that this object takes an extremely simple form
in the gauge $A^+=0$, especially when compared to what one obtains in
the (covariant) Lorenz gauge (see appendix \ref{sec:lorenz}). This
transfer matrix is central in deriving the solution of Yang-Mills
equations for pA collisions, and will be a crucial building block for
calculating higher order corrections in the weak source to this
solution.

\section*{Acknowledgements}
We would like to thank R. Baier, I. Balitsky, J.-P. Blaizot, D. Dietrich, E. Iancu, A.H. Mueller,
D. Schiff, and R. Venugopalan for useful discussions on the issues
discussed in this paper.

\appendix

\section{Gluon propagator in covariant gauge}
\label{sec:lorenz}
It is sometimes useful to have expressions for the propagation of a
gluon field inside the nucleus in a covariant gauge ($\partial_\mu
A^\mu=0$). We derive in this appendix the analogue of
eqs.~(\ref{eq:LC-prop}) for the Lorenz gauge. In this gauge, the color
field of the nucleus is the same as its field in the $A^+=0$ gauge:
\begin{equation}
A_0^+=A_0^i=0\quad,\quad A_0^-=-g\delta(x^+)\frac{1}{{\bs\partial}_\perp^2}\rho(\x_\perp)\; ,
\end{equation}
thanks to the independence of the nuclear sources on $x^-$. The
Yang-Mills equation that controls the evolution of $A_1^+$ is very
simple and reads
\begin{equation}
\square A_1^+-ig[A_0^-,\partial^+A_1^+]\; ,
\end{equation}
and its solution can be written as
\begin{equation}
A_1^+(\epsilon)=V A_1^+(0)\; ,
\label{eq:A+-lorenz-prop}
\end{equation}
where we use the same compact notations as in eq.~(\ref{eq:LC-prop})
and where $V$ is a Wilson line,
\begin{equation}
V\equiv {\cal T}_+ \,\exp \left[i \frac{g}{2}\int_0^\epsilon
dz^+\; A_{0a}^-(z^+,\x_\perp)T^a\right]\; ,
\end{equation}
that differs from $U$ only in the factor $1/2$ in the exponential.

The Yang-Mills equation for $A_1^i$ reads
\begin{eqnarray}
\square A_1^i-2ig[A_0^-,\partial^+A_1^i]
=ig (\partial^i A_0^-\cdot T) A_1^+-ig(A_0^-\cdot T)\partial^i A_1^+\; .
\end{eqnarray}
The solution of this equation reads
\begin{equation}
A_1^i(\epsilon)=U A_1^i(0)
+
i\frac{g}{2} U\otimes(\partial^iA_0^-\cdot T)\otimes V
\frac{1}{\partial^+}A_1^+(0)
-
i\frac{g}{2} U\otimes(A_0^-\cdot T)\otimes \frac{\partial^i}{\partial^+}(VA_1^+(0))\; ,
\label{eq:Ai-Lorenz-1}
\end{equation}
where we have used the following compact notation:
\begin{equation}
U_1\otimes A \otimes U_2 \equiv \int_0^\epsilon dx^+\;
U_1(\epsilon,x^+,\x_\perp){\cal A}(x^+,\x_\perp)U_2(x^+,0,\x_\perp)\;
,
\end{equation}
with $U_{1,2}$ any pair of Wilson lines evaluated at the same
transverse coordinate, and ${\cal A}$ the nuclear field or one of its
derivatives. With this notation, one has
\begin{eqnarray}
&&
\partial^i U=ig U\otimes (\partial^iA_0^-\cdot T)\otimes U\; ,
\nonumber\\
&&
\partial^i V = i\frac{g}{2} V\otimes (\partial^iA_0^-\cdot T)\otimes V\; ,
\label{eq:partialUV}
\end{eqnarray}
and the slightly less obvious relation
\begin{equation}
U-V=i\frac{g}{2}U\otimes(A_0^-\cdot T)\otimes V
=i\frac{g}{2}V\otimes(A_0^-\cdot T)\otimes U\; .
\label{eq:U-V}
\end{equation}
A proof of the latter formula was given in \cite{BlaizGV1}. Thanks to
these relations, it is straightforward to simplify
eq.~(\ref{eq:Ai-Lorenz-1}) into
\begin{equation}
A_1^i(\epsilon)=U A_1^i(0)
+\partial^i\Big[V\frac{1}{\partial^+}A_1^+(0)\Big]
-U\frac{\partial^i}{\partial^+}A_1^+(0)\; .
\label{eq:Ai-lorenz-prop}
\end{equation}
One can see that all the convolutions between $U$'s and $V$'s have
disappeared from this expression.

Let us consider finally the Yang-Mills equation for $A_1^-$. This
equation involves the current $J^-$, and in the Lorenz gauge the field
$A_1^+$ induces a color precession of $J^-$, which can be seen as a
correction $J_1^-$ of order ${\cal O}(A_1^\mu)$ to this component of
the current. The equation that determines this correction is
determined from the covariant current conservation, which can be
satisfied by
\begin{eqnarray}
&& J_1^+=J_1^i=0\; ,
\nonumber\\
&&\partial^+ J_1^- = ig [A_1^+,J_0^-]\; ,
\end{eqnarray}
where $J_0^-$ is the minus component of the nuclear current in the
absence of any extra field. If we recall that
$J_0^-=-{\bs\partial}_\perp^2 A_0^-$, we can solve the last
equation as
\begin{equation}
J_1^-=ig({\bs\partial}_\perp^2 A_0^-\cdot T) V \frac{1}{\partial^+} A_1^+(0)\; .
\end{equation}
The Yang-Mills equation for $A_1^-$ reads
\begin{eqnarray}
\square A_1^- -ig[A_0^-,\partial^+A_1^-]
&=& J_1^-
+(ig)^2 (A_0^-\cdot T)^2 A_1^+
\nonumber\\
&&
-2ig (A_0^-\cdot T) \partial^- A_1^+ -ig (\partial^- A_0^-\cdot T) A_1^+
\nonumber\\
&&
+2ig (\partial^i A_0^-\cdot T) A_1^i +ig (A_0^-\cdot T) \partial^i A_1^i\; .
\end{eqnarray}
It is straightforward to solve this equation and write its solution as
\begin{eqnarray}
&&
A_1^-(\epsilon)=
VA_1^-(0)
\nonumber\\
&&\;\; + i\frac{g}{2} V\otimes ({\bs\partial}_\perp^2A_0^-\cdot
T)\otimes V \frac{1}{\partial^{+2}}A_1^+(0) +\frac{(ig)^2}{2}
V\otimes (A_0^-\cdot T)^2 \otimes V \frac{1}{\partial^+} A_1^+(0)
\nonumber\\
&&\;\;
-ig V\otimes (A_0^-\cdot T)\otimes \partial^-
\Big[V\frac{1}{\partial^+}A_1^+(0)\Big]
-i\frac{g}{2} V\otimes (\partial^-A_0^-\cdot T)\otimes V\frac{1}{\partial^+}
 A_1^+(0)
\nonumber\\
&&\;\;
+ig V\otimes (\partial^i A_0^-\cdot T) \otimes
\Big[U\frac{1}{\partial^+} A_1^i(0)+\partial^i\Big(V\frac{1}{\partial^{+2}}A_1^+(0)\Big)-U\frac{\partial^i}{\partial^{+2}}A_1^+(0)\Big]
\nonumber\\
&&\;\;
-i\frac{g}{2} V\otimes (A_0^-\cdot T)\otimes \partial^i\Big[
U\frac{1}{\partial^+} A_1^i(0)+\partial^i\Big(V\frac{1}{\partial^{+2}}A_1^+(0)\Big)-U\frac{\partial^i}{\partial^{+2}}A_1^+(0)
\Big]
\; .
\nonumber\\
&&
\end{eqnarray}
After some lengthy manipulations based on eqs.~(\ref{eq:partialUV})
and (\ref{eq:U-V}), one can eliminate all the convolution products
$\otimes$ and obtain
\begin{eqnarray}
&&
A_1^-(\epsilon)=
VA_1^-(0)
+\Big[\partial^iU\frac{1}{\partial^+} 
-V\frac{\partial^i}{\partial^+}\Big]A_1^i(0)\nonumber\\
&&
\quad\!\!\!\!+\Big[\frac{1}{2} ({\bs\partial}_\perp^2 U)
\frac{1}{\partial^{+2}}-\frac{1}{2}
U\frac{{\bs\partial}_\perp^2}{\partial^{+2}}-\frac{1}{2}
{\bs\partial}_\perp^2U\frac{1}{\partial^{+2}}
-(\partial^-V)\frac{1}{\partial^+}+{\bs\partial}_\perp^2V\frac{1}{\partial^{+2}}\Big]A_1^+(0)\; .
\nonumber\\
&&
\label{eq:A--lorenz-prop}
\end{eqnarray}
Eqs.~(\ref{eq:A+-lorenz-prop}), (\ref{eq:Ai-lorenz-prop}) and
(\ref{eq:A--lorenz-prop}) thus constitute the equivalent of
eqs.~(\ref{eq:LC-prop}) in the Lorenz gauge. One can see in these
expressions the advantage of working in the light-cone gauge: not only
the formulas as much more compact, but in addition they do involve
only the Wilson line $U$, and not $V$. As a side note, one can check that
\begin{equation}
\partial_\mu A_1^\mu(\epsilon)= V \partial_\mu A_1^\mu(0)\; ,
\end{equation}
i.e. the Lorenz gauge condition is satisfied at $x^+=\epsilon$
provided that it was satisfied by the field incoming at $x^+=0$.

\bibliographystyle{unsrt}

\end{document}